\documentclass[12pt]{article}

\usepackage{epsfig}

\usepackage{amsmath}
\usepackage{amssymb}
\usepackage{euscript}

\newcommand{\veps}{\varepsilon}

\newcommand{\tH}{{\hat{t}}}

\begin{document}

\begin{titlepage}


\begin{flushright}
\bf IFJPAN-V-05-09
\\
\bf CERN-PH-TH/2005-146
\end{flushright}

\vspace{1mm}
\begin{center}
  {\Large\bf%
    Constrained non-Markovian Monte Carlo modelling of the evolution
    equation in QCD$^{\star}$
}
\end{center}
\vspace{3mm}

\begin{center}
{\bf S. Jadach}
{\em and}
{\bf M. Skrzypek} \\

\vspace{1mm}
{\em Institute of Nuclear Physics, Academy of Sciences,\\
  ul. Radzikowskiego 152, 31-342 Cracow, Poland,}\\
and\\
{\em CERN Department of Physics, Theory Division\\
CH-1211 Geneva 23, Switzerland}
\end{center}

\vspace{5mm}
\begin{abstract}
A new class of the constrained Monte Carlo (CMC) algorithms
for the QCD evolution equation was recently discovered.
The constraint is imposed on the type and the total longitudinal
energy of the parton exiting QCD evolution and entering a hard process.
The efficiency of the new CMCs is found to be reasonable.
\end{abstract}

\vspace{4mm}
\begin{center}
\em Contribution to the HERA--LHC Workshop, 2004--2005
\end{center}

\vspace{5mm}
\begin{flushleft}
{\bf IFJPAN-V-05-09
\\
\bf CERN-PH-TH/2005-146
}
\end{flushleft}

\vspace{3mm}
\footnoterule
\noindent
{\footnotesize
$^{\star}$Supported in part by the EU grant MTKD-CT-2004-510126,
  in partnership with the CERN Physics Department.
}

\end{titlepage}

\section{Introduction}
This brief report summarizes the recent developments in the area
of the Monte Carlo (MC) techniques for the perturbative QCD calculations.
Most of it was done at the time of the present HERA--LHC workshop,
partial results being presented at several of its meetings.
At present, two papers, \cite{Jadach:2005yq} and \cite{Jadach:2005bf},
demonstrating the principal results are already available.
Generally, these MC techniques concern the QCD evolution of the parton
distribution functions (PDFs) $D_k(x,Q)$,
where $k$ denotes the type of the parton (quark, gluon),
$x$ the fraction of longitudinal momentum  of the initial hadron 
carried by the parton,
and the size of the available real/virtual emission phase space is $Q$.
The evolution equation describes the response of the PDF to an increase
of $Q$; $D_k(x,Q)$ is an inclusive distribution and can be measured
almost directly in 
hadron--lepton scattering.
On the other hand, it was always known
that there exists in QCD an {\em exclusive} picture of the PDF,
the so-called parton-shower process,
in which $D_k(x,Q)$ is the distribution of the parton exiting the emission
chain and entering the hard process (lepton--quark for example).
The kernel functions $P_{kj}(Q,z)$,
that govern the differential evolution equations of PDFs
are closely related to distributions governing a single emission
process $(i-1)\to i$ in the parton shower: $P_{k_i k_{i-1}}(Q_i,x_i/x_{i-1})$.

In other words, the evolution ($Q$-dependence) of PDFs
and the parton shower represent two faces of the same QCD reality.
The first one (inclusive) is well suited for basic precision tests of QCD
at hadron--lepton colliders,
while the second one (exclusive) provides realistic exclusive Monte Carlo modelling,
vitally needed for experiments at high-energy particle colliders.

At this point, it is worth stressing that, so far, we were referring 
to DGLAP-type PDFs~\cite{DGLAP} and their evolution,
and to constructing a parton-shower MC starting from them,
as was done two decades ago and is still done today.
This involves a certain amount of ``backward engineering'' and educated guesses,
because the classical inclusive PDFs integrate over the $p_T$ of the exiting parton.
The so-called unintegrated PDFs (UPDFs) 
$D_k(x,p_T,Q)$ would be more suitable for the purpose,
leading to higher-quality QCD calculations.
UPDFs are, however, more complicated to handle, both numerically and theoretically.
(It is still a challenge to construct a parton-shower MC 
based consistently on the theoretically well defined UPDFs.)

Another interesting ``entanglement'' of the evolution of PDFs on one side
and of the parton shower (PS) MC on the other side
is also present in the modelling of the showering of the incoming hadron --
mostly for technical reasons and convenience.
The Markovian nature of the QCD evolution can be exploited directly in the PS MC,
where partons split/decay as long as there is enough energy to dissipate (final state)
or the upper boundary $Q$ of the phase space is hit (initial state).
The multiparton distribution in such a MC is a product of the evolution kernels.
However, such a direct Markovian MC simulation of a shower is hopelessly
inefficient in the initial state, because the hard process accepts only
certain types and momenta of the incoming partons --
most of the {\em shower histories} are rejected (zero MC weight) by the hard process.
This is especially true for forming narrow resonances
such as electroweak bosons or Higgs boson at the LHC.
The well known ``workaround'' is the backward evolution MC algorithm of Sj\"ostrand,
used currently in all PS MCs, such as
HERWIG~\cite{Corcella:2000bw} and PYTHIA~\cite{Sjostrand:2000wi}.
Contrary to the forward Markovian MC, where the physics inputs are PDFs
at low $Q_0\sim$1 GeV
and the evolution kernels, in the backward evolution MC one has to know
PDFs in the entire range $(Q_0,Q)$ from a separate non-MC numerical program
solving the evolution equation and setting up 
look-up tables
(or numerical parametrization) for them%
\footnote{Backward evolution 
is basically a change in the order of the generation of the variables:
Consider generating $\rho(x,y)$, where one generates
first $x$ according to $\rho(x)=\int dy\; \rho(x,y)$,
and next $y$ according to $\rho(x,y)$,
by means of {\em analytical} mappings of $x$ and $y$ into uniform random numbers.
However, such analytical mappings may not exist,
if we insist on generating first $x$ and next $y$!
Nevertheless, we may still proceed with the same method by ``brute force'',
if we pretabulate and invert numerically the functions
$R(x)=\int^x \int dx' dy'\; \rho(x',y')$ and $R_x(y)=\int^y dy'\; \rho(x,y')$.
This is what is done in a more dimensional case of the backward-evolution MC;
it also explains why pretabulated PDFs are needed in these methods.
}.

The following question has been pending
in the parton-shower MC methodology for a long time:
Could one invent an efficient ``monolithic'' MC algorithm for the parton shower
from the incoming hadron, in which no external PDFs are needed and the
only input 
are PDFs at $Q_0$ and the evolution kernel
(the QCD evolution being a built-in feature of the parton shower MC)?
Another question rises immediately: Why bother? 
Especially since this is a tough technical problem.
This cannot still be fully answered before the above technique is applied
in the full-scale (four-momentum level) PS MC.
Generally, we hope that this technique will open new avenues in the development
of the PS MC at the next-to-leading-logarithmic (NLL) level.
In particular, it may help in constructing PS MCs closely related
to unintegrated structure functions and, secondly,
it may provide a better integration of the NLL parton shower
(yet to be implemented!)
with the NLL calculation for the hard process.

The first solution of the above problem of finding an efficient
``constrained MC'' (CMC) algorithm for the QCD evolution was presented
in refs.~\cite{Jadach:2005yq,zinnowitz04}.
This solution belongs to what we call a CMC class II, and it relies on
the observation that all initial PDFs at $Q_0$ can be approximated
by $\hbox{const} \cdot x_0^{\eta -1}$;
this is to be corrected by the MC weight at a later stage.
This allows elimination of the constraint $x=\prod_i z_i$,
at the expense of $x_0$, keeping the factorized form of the products
of the kernels.
Simplifying phase-space boundaries in the space of $z_i$ is the next
ingredient of the algorithm.
Finally, in order to reach a reasonable MC efficiency for the pure
bremsstrahlung case out of the gluon emission line,
one has to generate a $1/z$ singularity in the
$G\to G$ kernel in a separate branch of the MC.
The overall efficiency of the MC is satisfactory, as is demonstrated
in ref.~\cite{Jadach:2005yq} for the case of 
the pure bremsstrahlung out of the gluon and quark colour charge.
Generalization to the quark--gluon 
transition is outlined, but not yet implemented.
The main drawback of this method is its algebraic complexity.
Further improvement of its relatively low MC efficiency is possible
(even though it could lead to even more algebraic complexity).

The second, more efficient, CMC algorithm was presented
in ref.~\cite{Jadach:2005bf}
(as well as during the October 2004 meeting of the workshop).
It belongs to what we call a CMC class I.
The main idea is to project/map points from
the hyperspace defined by the energy constraint $x=\prod_i z_i$, into
a simpler hyperspace, defined by the hardest emission, $x= \min z_i$.
This mapping is accompanied by the appropriate MC weight,
which compensates exactly for the deformation of the distributions involved,
and the bookkeeping of the hyperspace boundaries is rigorous.
The above describes a CMC for the pure bremsstrahlung segment of the
gluon emission out of a quark or gluon chain.
Many such segments are interconnected by the quark--gluon transitions.
The algebraic hierarchic reorganization of the emission chain into a 
super-level
of the quark--gluon transitions and sub-level of the pure bremsstrahlung
is an important ingredient in all CMC algorithms
and will be published separately \cite{raport04-09}.
The basic observation made in ref.~\cite{Jadach:2003bu} is that
the average number of super-level transitions is low, $\sim 1$;
hence for precision of a $10^{-4}$ it is sufficient to
limit it to three or four transitions.
The integration/simulation of the super-level variables is done efficiently
using the general-purpose MC tool FOAM \cite{foam:2002,Jadach:2005ex}.
The above proof of the correctness
of the CMC class I algorithm concept was given in ref.~\cite{Jadach:2005bf}
for the full DGLAP-type QCD evolution with the LL kernels
(including quark--gluon transitions).

Although our main aim is to construct
the non-Markovian CMC class of algorithms,
we have developed in parallel the family of Markovian MC (MMC) 
algorithms/programs, which provide numerical solutions of the QCD evolution equations
with high precision, $\sim10^{-3}$.
We use them at each step of the CMC development as numerical benchmarks
for the precision tests of the algorithms and their software implementations.
The first example of MMC for DGLAP at LL was defined/examined
in ref.~\cite{Jadach:2003bu} and tested using the non-MC
program QCDnum16~\cite{qcdnum16}%
\footnote{
It was also compared with the non-MC program APCheb \cite{APCheb33}.
}.
In some cases our MMC programs stand ahead of their CMC brothers;
for instance, they already include NLL DGLAP kernels.
A systematic description of the MMC family of our MC toolbox is
still under preparation \cite{raport05-03}.

The last development at the time of the workshop
was an extension of the CMC type-I algorithm from DGLAP
to CCFM one-loop evolution~\cite{CCFM}
(also referred to as HERWIG evolution \cite{Nason:2004rx}),
in which the strong coupling constant gains $z$-dependence,
$\alpha_s(Q)\to \alpha_s(Q(1-z))$, as advocated in ref.~\cite{Amati:1980ch},
confirmed by NLL calculations \cite{Curci:1980uw}.
The above ansatz also compels introduction of a $Q$-dependent IR cutoff,
$\veps=Q_\veps/Q$: another departure from DGLAP.
This version of the CMC is still unpublished. 
Its version for the pure bremsstrahlung was presented at the March 2005
meeting of the workshop;
in particular a perfect numerical agreement with the couterpartner MMC
was demonstrated.
Recently both CMC and MMC for the one-loop CCFM were 
extended to quark--gluon transitions, and again perfect agreement was found.

\begin{figure}[!ht]
  \centering
  {\epsfig{file=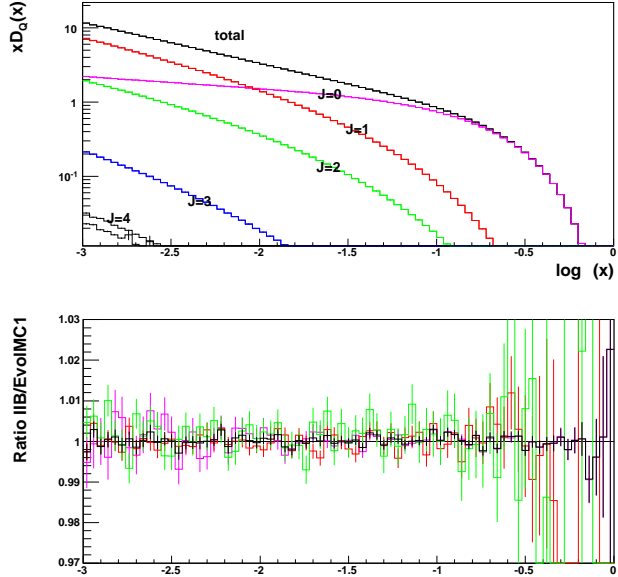, width=90mm}}
  \caption{\sf
    CMC of the one-loop CCFM versus the corresponding MMC for quarks; 
    number of quark--gluon transitions $J=0,1,2,3,4$, and the total.
    The ratios in the lower plot are for $n=0,1$ and the total (blue).
    }
  \label{fig:CMCHWdlnx}
\end{figure}

For the detailed description of the new CMC algorithm, we refer
the reader to the corresponding papers
\cite{Jadach:2005yq} and \cite{Jadach:2005bf} and workshop presentations%
\footnote{To be found at {\tt http://jadach.home.cern.ch/jadach/}.}.
Here, let us only show one essential step in the development of the
CMC for the one-loop CCFM model --
the mapping of the Sudakov variables for the pure bremsstrahlung:
\begin{equation}
\begin{split}
I&=\int_{t_0}^{t_1} dt \int_{0}^{z_1} dz\; \alpha(Q(1-z))\; zP^\Theta_{GG}(z,t)\\
& =\frac{2}{\beta_0} \int_{0}^{z_1} dz \int_{t_0}^{t_1} dt\;
       \frac{1}{\tH +\ln(1-z)} \frac{\theta_{\ln(1-z)>\tH_\veps-\tH}}{1-z}
  =\frac{2}{\beta_0} \int_0^{y_{\max}} dy(z) \int_0^1 ds(t).
\end{split}
\end{equation}
The short-hand notation 
$\tH = \tH(t) \equiv t-t_\Lambda$ and $v=\ln(1-z)$ supplements
that of ref.~\cite{Jadach:2005bf} in use, and the mapping reads
\begin{equation}
\begin{split}
  &y(z)=\rho(v_1;\tH_1,\tH_0) 
     = \rho(v_1+\tH_1) -\theta_{v_1>t_\veps-t_0} \rho(v_1+\tH_0),
\quad
  s(t)= \frac{\ln(\tH +v)}{\rho'(v;\tH_1,\tH_0)},\qquad
\\&
 \rho'(v;\tH_1,\tH_0)=
    \theta_{v<t_\veps-t_0} \rho'(v+\tH_1)
   +\theta_{v>t_\veps-t_0} [\rho'(v+\tH_1)-\rho'(v+\tH_0)],
\end{split}
\end{equation}
where $\rho(t)\equiv \tH(\ln\tH-\ln\tH_\veps)+\tH_\veps-\tH$.
Once the above mapping is set, the same algorithm, with the parallel
shift $y_i\to y_i+Y$, can be used in this case.
The super-level of quark--gluon transitions is again implemented using FOAM%
\footnote{The  $z$-independent $\alpha_s(t)$ is set in front
of the relevant flavour-changing kernels to simplify the program.}.
A numerical comparison of the corresponding CMC and MMC programs
is shown in fig.~\ref{fig:CMCHWdlnx}. 
The MC efficiency is comparable with that of the DGLAP case.

{\em Summary:}
We have constructed and tested new, efficient, constrained MC algorithms
for the initial-state parton-emission process in QCD.

\providecommand{\href}[2]{#2}

\end{document}